\documentclass[a4paper,10pt]{article}
	\usepackage{enumerate}
	\usepackage{color}
	\usepackage[utf8]{inputenc} 
	\usepackage[english]{babel}
	\usepackage[T1]{fontenc}
	\usepackage{graphicx}
	\usepackage{amsfonts,amssymb,amsmath,latexsym,amsthm}
	\usepackage{textcomp}
	\usepackage[pdftex]{hyperref}
	\usepackage{geometry}
	\usepackage{tikz}
	\newcommand{\C} {\mathcal{C}}
	\newcommand{\X}{\mathbf{ X}}
	\newcommand{\Y}{\mathbf{ Y}}

	\author{ M. V. Flamarion$^1$,   E. Kochurin$^{2,3}$, R. Ribeiro-Jr$^4$,  N. Zubarev$^{2,5}$}
	\title{Flow structure beneath periodic waves with constant vorticity  under strong horizontal  electric fields }

	\date{}
	
\begin{document}
	\maketitle
	\begin{center}

		{\footnotesize $^1$ Department of Mathematics, Aeronautics Institute of Technology (ITA), São José dos Campos, SP 12228-900, Brazil \\
	marcelovflamarion@gmail.com }

		\vspace{0.3cm}

		{\footnotesize $^2$UInstitute of Electrophysics, Ural Branch of Russian Academy of Sciences, Yekaterinburg, 620016, Russia\\
		$^3$Skolkovo Institute of Science and Technology, Moscow, 121205, Russia\\
		kochurin@iep.uran.ru }
	
	\vspace{0.3cm}

	{\footnotesize $^{4}$ UFPR/Federal University of Paran\'a,  Departamento de Matem\'atica, Centro Polit\'ecnico, Jardim das Am\'ericas, Caixa Postal 19081, Curitiba, PR, 81531-980, Brazil \\ robertoribeiro@ufpr.br}
	
	\vspace{0.3cm}
	
		{\footnotesize $^5$P. N. Lebedev Physical Institute, Russian Academy of Sciences, 119991 Moscow, Russia\\
		nick@iep.uran.ru }
	
	\vspace{0.3cm}
	

	\end{center}

	
	\begin{abstract} 
	\noindent 
	While several articles have been written on Electrohydrodynamics (EHD) flows or flows with constant vorticity separately, little is known about the extent to which the combined effects of EHD and constant vorticity affect the flow. This study aims to fill this gap by investigating how a horizontal electric field and constant vorticity jointly influence the free surface and the emergence of stagnation points. Using the Euler equations framework, we employ conformal mapping and pseudo-spectral numerical methods. Our findings reveal that increasing the electric field intensity eliminates stagnation points and  smoothen the  wave profile. This implies  that a horizontal electric field acts as a mechanism for the elimination of stagnation points within the  fluid body.
	


		\end{abstract}

		\section{Introduction}
	
	Electrohydrodynamics (EHD) flows and flows with constant vorticity are  two topics of interest to researchers from different areas.  EHD flows consist of the coupling between fluid and electromagnetism.  Its motivation comes from the industrial interest in controlling fluid motion trough electric field manipulation  \cite{HB, PS, CPP, E, MS,CCY}. This application extends to cooling systems in conducting pumps \cite{Ghoshal}, coating processes \cite{Griffing}, among others. Constant vorticity flows serve as a model for waves within linearly sheared currents, representing realistic flow scenarios in certain conditions as for instances in flows when waves are long compared with the depth \cite{Teles Da Silva}. 
		
		Most studies on EHD and flows with constant vorticity have typically examined these topics separately. 
		 The literature in those  topics is broad and therefore it is difficult to give a comprehensive overview of contributions. For the interested reader, we mention a few references from which the bibliography may be useful.

		 The paper by Papageorgiou \cite{Papageorgiu:2019} offers a good review on EHD flows, perfectly describing the effects of electric fields on immiscible film flows. Specifically, it highlights that normal electric fields have the capability to destabilize the interface between two fluids\cite{Ozen}, whereas tangential electric fields, conversely, can stabilize interfacial waves due to these effects in a way that   Rayleigh-Taylor\cite{CPP}  and Kelvin-Helmholtz\cite{Sayed,EFARF,ZK} instabilities  can be suppressed.

	Regarding flows with constant vorticity, one of their primary characteristics is the presence of overhanging waves and the emergence of stagnation points. Overhanging waves are free surface waves that are not a graph of a function. They have been demonstrated numerically \cite{VB96,VB94,DH19a,DH19b}, and their existence has also been rigorously proven for periodic waves \cite{CSV,HW22}. Stagnation points represent fluid particles that remain stationary in the wave moving frame. In irrotational flows, these points are charactersed by occuring at wave crests,  forming a sharp crest \cite{Varvaruca}.  By adjusting the vorticity in constant vorticity flows, stagnation points may emerge within the fluid bulk, creating  a recirculation zone whose profile resembles Kelvin's cat eye flow \cite{Ribeiro:2017,Vasan,Teles Da Silva,Flamarion:2023Eng}.


	All previously mentioned studies have primarily focused on  either  EHD flows or flows with constant vorticity.  Only recently research efforts have shifted towards investigating the behavior of electrohydrodynamic flows within the context of constant vorticity. This approach commenced with the work of Hunt and Dutykh 
\cite{HD} which concerned in  the derivation of linear and weakly nonlinear models for rotational waves in the presence of  vertical electric fields.  
The underlying flow structure beneath waves with constant vorticity under the influence of vertical electric fields has been investigated more recently by Flamarion and collaborators  \cite{Flamarion:2022PoF,FGR}. Their findings highlighted that by maintaining a fixed vorticity, the increase in the intensity of the electric field can lead to the appearance of stagnation points.  This phenomenon subsequently induces the formation of cat-eye structure.  This result underscored the potential for controlling stagnation points through the manipulation of electric field strength rather than adjusting vorticity. The first investigation into the combination of a horizontal 
electric field and linear shear flow was conducted by  Flamarion et al. \cite{Kochurin:2023} . Their work demonstrates that in the high vorticity regime in the absence of  an external electric field, the evolution of the free surface exhibits notably intricate nonlinear behavior. Specifically, observations include the occurrence of wave breaking resulting in the formation of an air bubble within the fluid. In contrast, the introduction of a horizontal electric field ensures the preservation of the free surface wave profile.

As far as we are aware, the effect of horizontal electric fields on the flow structure beneath a surface wave remains unexplored. The current study aims to contribute as an additional effort towards addressing this gap in the literature. In particular, we focus on numerically investigating the combined influence of a horizontal electric field and constant vorticity on the behavior of the free surface and the emergence of stagnation points. Our investigation is carried out within the framework of Euler equations, utilizing computation made through a conformal mapping technique and pseudo-spectral numerical methods. 
We find that  the increase of the intensity of the electric field leads to the elimination of stagnation points and tends to smoothen the wave profile.  This implies  that a horizontal electric field acts as a mechanism for the elimination of stagnation points within the fluid body.  This behaviour  is the opposite  of the effect of  normal electric field where an increase in electric  field stimulates the  emergence of stagnation points \cite{Flamarion:2022PoF}. 
		
This paper is organized as follows. Section 2 presents the mathematical formulation, while Section 3 provides a succinct overview of the conformal mapping technique and the numerical approach employed. Section 4 outlines the linear theory, followed by the presentation of results obtained from the full Euler equations in Section 5. Finally, Section 6 encompasses a comprehensive discussion of the results with a  summary of concluding remarks.

\section{Mathematical Formulation}

\begin{figure}[!ht]
		\centering
		\includegraphics[scale=1]{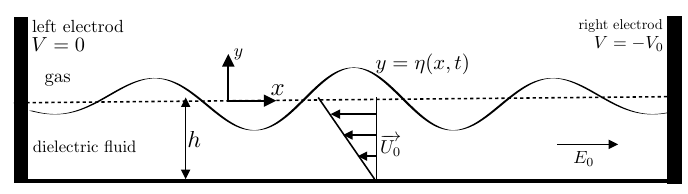}
				\caption{Schematic illustration of the geometry of the problem.}
		\label{fig:fig22}
	\end{figure}

Let us consider an inviscid and incompressible dielectric fluid (there are no  electric charges inside the fluid) with density $\rho$ and electric permittivity $\epsilon_1$ on channel of  mean depth $h$ in  a two-dimensional space. The fluid is bounded laterally by two wall electrode and at the top by a  non-conducting gas.    The  electric field acts horizontally  in the fluid, with a magnitude represented by $E_0=V_0/L$, where $L$ signifies the system horizontal size and $V_0$ represents the potential difference along the horizontal axis. In the electrostatic limit of Maxwell's equations, the induced magnetic fields are negligible, then we obtain an irrotational electric field due to Faraday's law. This is expressed via a potential function $V(x, y)$ governing the electric field $\vec{E}$, such that $\vec{E}=-\nabla V$. Consequently, $V$ satisfies the Laplace Equation within the dielectric fluid layer. The schematic representation of this problem is depicted in  Figure \ref{fig:fig22}. Establishing a Cartesian coordinate system $\{x,y\}$ with gravity oriented in the negative $y$-direction and $y=0$ denoting the undisturbed free surface, the unperturbed state reveals the electric field potential to adopt a linear form: $V(x, y)=-(V_0/L)x$. The fluid lower boundary resides at $y=-h$, while the upper boundary, labeled $\eta(x, t)$, remains free to move.

We consider  that the  flow in the dieletric fluid is rotational  with constant vorticity  $\gamma$. In the unperturbed state, the velocity field in the fluid body is given by $$\vec{U}_0 = \left(-\gamma(y+h),0\right).$$
The background flow $\vec{U}_0$ is a linear shear flow with horizontal velocity equal to zero at the bottom.  Let us consider  $\phi$ as an potential velocity of an  irrotational perturbation  of the shear flow, i.e,  the velocity field in the fluid body is given as 
\begin{equation}\label{eq:velocity_field}
	\vec{U} = \vec{U}_0 +\nabla \phi.
\end{equation}

The goal of this work is to identify steady waves with wavelength  $L$ in a frame of reference that moves with the wave speed $c$. This is achieved by introducing the change of variables $X=x-ct$, $Y = y$, which yields $\eta(x,t)=\eta(X)$.  In this context, the dimensionless  governing equation can be written in terms of the potential $\phi$ and the voltage potential $V$ as follows

\begin{align}
	\nabla^2 \phi &= 0 & \text{in} -1<Y<\eta(X), \label{eq:field1}\\
	\nabla^2 V &= 0 & \text{in} -1<Y<\eta(X), \label{eq:field2} \\
		-c \eta_X + \left(\phi_X - \gamma\left(\eta + 1 \right) \right) \eta_X &=\phi_Y & \text{for} \quad   Y=\eta(X), \label{eq:KBC2} \\
	\phi_Y &=0 & \text{for} \quad  Y=-1, \label{eq:KBC1}\\
	\partial V/\partial n&=0 & \text{for} \quad Y=\eta(X), \label{eq:KBCV1}\\
	V_y&=0 & \text{for}  \quad  Y=-1, \label{eq:KBCV2}
\end{align}
where $\partial/\partial n$  denotes the normal derivative at the  free surface $\eta(x,t)$.
Additionally, the dynamic boundary condition is obtained by ensuring the continuity of pressure on the free surface, leading to the following expression
\begin{equation}\label{eq:Bern1}
	-c\phi_X + \frac{1}{2} (\phi_X^2 + \phi_Y^2)  - \gamma \left(\eta + 1\right) \phi_X + \gamma \psi -   \frac{E_b}{2}\lvert \nabla V \rvert^2=B, 
\end{equation}
 where $\psi$ is the harmonic conjugate of $\phi$,   $E_b={\epsilon_1 V_0^2}/\rho g h^3$   is the nondimensional parameter characterizing the intensity of the  electric field and $B$ is the Bernoulli constant.     Here, we consider the case of a strong electric field, i.e., the effects of gravity and capillarity are not taken into account. Besides,  we highlight that this set of equations is written in the dimensionless form where we have chosen  $h$, $h/g$ and $V_0$ as the reference length, time and voltage potential, respectively.  
 
 The stream function can be written as 
 \begin{equation}\label{eq:streamfunction}
  \psi_s(X,Y) = \psi(X,Y) - \gamma Y\left( \frac{Y}{2} + 1\right)  - cY.
  \end{equation}
 
 In the scenario of a fluid with a high dielectric constant, the electric field lines are align tangentially to the  free surface, indicating a tangential field orientation  \cite{Zubarev:2004,Zubarev:2009}. Consequently, the electrostatic matter can be entirely resolved using the method of conformal transformation,  a topic that will be elaborated upon in the subsequent section.

\section{Conformal mapping and numerical method}

The present study employs a pseudo-spectral numerical method  to solve the equations (\ref{eq:field1})-(\ref{eq:Bern1}). The chosen approach is based on conformal mapping, which facilitates the transformation of the physical domain into a canonical one with simpler geometry. This technique has been previously applied in the study of rotational periodic waves \cite{Ribeiro:2017}, as well as rotational  solitary  waves \cite{ Flamarion:2023Eng}. Addionally, it has been employed  to  investigate the effect of a normal electric field on the flow structure beneath periodic rotational waves  \cite{Flamarion:2022PoF} and, more recently,  in examining  the effects of external horizontal electric fields on the propagation of free surface waves in shear flow\cite{Kochurin:2023}. 
Essentially, the physical domain is mapped onto a horizontal strip of uniform thickness denoted as $D$, and all computations are carried out within this canonical domain. A concise overview of the numerical scheme is provided here.

Let us denote by $\{\xi,\zeta\}$ the coordinate system in the canonical domain and by $X(\xi,\zeta)$ and $Y(\xi,\zeta)$ the horizontal and vertical component of the conformal map under which a strip of width $D$ is mapped onto the physical domain.  By elementary computation it is easy to find that 
\begin{equation}\label{eq:Y0}
Y(\xi,\zeta) = \mathbf{F}^{-1} \left[  \frac{\sinh(k_j(\zeta+D))}{\sinh(k_jD)}  \mathbf{F}_{k_j}[\mathbf{Y}]  \right] + \frac{\zeta}{D},  
\end{equation}
and
\begin{equation}\label{eq:X0}
X(\xi,\zeta) =  \xi - \mathbf{F}^{-1} \left[i  \frac{\cosh(k_j(\zeta+D))}{\sinh(k_jD)}  \mathbf{F}_{k_j}[\mathbf{Y}]  \right], \quad j \neq 0 
\end{equation}
where $ Y(\xi,0) = \Y(\xi) \equiv \eta(X(\xi,0),0)$, $k_j = 2\pi j/L$, for $j\in \mathbb{Z}$, and
$$ \mathbf{F}_{k_j}[g(\xi)] = \hat{g}(k_j) =  \frac{1}{L}\int_{-L/2}^{L/2}g(\xi) e^{-ik_j\xi}\, d\xi, $$
$$  \mathbf{F}^{-1}\left[ \{\hat{g}(k_j)\}_{j \in \mathbf{Z}}\right] = \sum_{j = -\infty}^{+\infty} \hat{g}(k_j) e^{ik_j\xi}.$$

Note that the profile of the free surface wave   in the physical domain is determined  by the trace of the curve  $(\X(\xi),\Y(\xi)) $, where $\X(\xi):=X(\xi,0)$ and $\Y(\xi):=Y(\xi,0)$.  Besides, we obtain  from  equation  (\ref{eq:Y0}) and (\ref{eq:X0}) that
\begin{equation}\label{eq:X_xi}
\X(\xi) =  \xi - \C[\Y(\xi)],
\end{equation}
with $\C[\cdot]:=\mathbf{F^{-1}}\left[ i\coth(k_jD)\mathbf{F}_{k_j}\left[\cdot\right] \right]$.

We choose the the wavelength in both  physical and canonical domain to be equal to $L$. This yields the equation
\begin{equation}\label{eq:D}
D = \left< \mathbf{Y} \right> + 1,
\end{equation}
where 
\begin{equation}
\left< \mathbf{Y} \right> =  \frac{1}{L}\int_{-L/2}^{L/2} \mathbf{Y} (\xi)\, d\xi.
\end{equation}

Upon establishing the conformal mapping formulas, the governing equations (\ref{eq:field1})-(\ref{eq:Bern1}) are reformulated in canonical coordinates. Subsequent manipulations are applied, resulting in the derivation of a single  governing equation for the free surface wave presented below
\begin{equation}\label{eq:Free surface}
	\begin{split}
		-\frac{c^2}{2} &- \frac{c^2}{2J}  +  \gamma^2 \frac{ (\C[(\mathbf{Y}+1)\mathbf{Y}_\xi])^2 - \left((\mathbf{Y}+1)\mathbf{Y}_\xi \right)^2}{2J} + \gamma^2\mathbf{Y}\left(\frac{\mathbf{Y}}{2} + b\right) \\
		&+ \frac{\left(c+\gamma \C [(\mathbf{Y}+1)\mathbf{Y}_\xi] \right) \left( c+ \gamma(\mathbf{Y}+1)(1 - \C[\mathbf{Y}_{\xi}]) \right)}{J} -\gamma c     - \frac{E_b}{2J} = B, 
	\end{split}
\end{equation}
where $J= \X_\xi^2  + \Y_\xi^2 $ is the Jacobian evaluated at $\zeta = 0$,  and $\X_\xi$ is the $\xi$-derivative of $\X(\xi)$, which is given by
$\X_\xi =  1- \C[\Y_\xi)]$.
It should be noted that in the absence of the shear flow, $\gamma=0$, the equation (\ref{eq:Free surface}) has exact solution for any shape of the surface profile: $c^2=E_b$, $B=-c^2/2$. In such fully electrohydrodynamic regime of motion, the fluid moves along the electric field lines. We refer to the work of Ribeiro-Jr et al. \cite{Ribeiro:2017} and Flamarion et al.  \cite{Kochurin:2023}  respectively for further details regarding the derivation of the rotational and electric components associated with equation (\ref{eq:Free surface}).

The free surface wave is assumed to be an even function with crest at $X=0$. Therefore, we consider that the crest is located at $\xi = 0$. We fix the $H$ height of  the free-surface elevation $\mathbf{Y}(\xi) = Y(\xi,0)$ through the equation
\begin{equation}\label{eq:height}
	\Y(0)- \Y(-L/2) = H.
\end{equation}
 To ensure that the mean level of the free surface wave is zero in the physical domain, we set
\begin{equation}\label{eq:mean0}
	\int_{-L/2}^{0} \Y\X_\xi , d\xi =0.
\end{equation}
Here, $[-L/2,L/2)$ is the computational domain, and recall that  $L$ is chosen such that the wavelength in the physical and canonical domain are the same. 

The system of equations comprising (\ref{eq:D}), (\ref{eq:Free surface}), (\ref{eq:height}) and (\ref{eq:mean0}) constitutes four equations with four unknowns: $\Y$, $c$, $D$, and $B$. The system is discretized in $\xi$ by introducing collocation points uniformly distributed along $\xi$. By evaluating all the $\xi$-derivatives via a Fourier spectral method using the FFT, the partial differential equations are transformed into algebraic equations, which can be solved using Newton's method, as detailed in \cite{Ribeiro:2017,Philo}.

	\section{Results from the Linear Theory}\label{ln}
	In this section, we will develop a linear theory based on the governing equations. We begin by considering a trivial solution characterized by
	
	\begin{equation}
		\phi_0(X,Y) = 0, \hspace{0.6cm} V_0(X,Y) = X, \hspace{0.6cm} \eta_0(X)=0.
	\end{equation}
	This solution is then perturbed by a small disturbance given by
	\begin{eqnarray}
		\begin{cases}\label{eq:linear1}
			\eta(X)=\epsilon \hat{\eta}, \
			\phi(X,Y)=\epsilon \hat{\phi}, \
			V(X,Y)= X+\epsilon \hat{V},
		\end{cases}
	\end{eqnarray}
	where $\epsilon$ is a small parameter that measures the wave amplitude.  By solving the Laplace equations \eqref{eq:field1} and \eqref{eq:field2}, along with the boundary conditions \eqref{eq:KBC1} and \eqref{eq:KBCV2}, we obtain the following expressions
	\begin{eqnarray}
			\begin{cases}\label{eq:linear2}
			\hat{\eta}(X)= \Re\left\{ A e^{ikX}\right \}\,, \\
			\hat{\phi}(X,Y)=\Re \left\{ M e^{ikX}\cosh k(Y+1)\right \}\,, \\
			\hat{V}(X,Y)= \Re\left\{  Ne^{ikX}\cosh k(Y+1)\right \}\,,
					\end{cases}
	\end{eqnarray}
	where $A$, $M$, and $N$ are unknown constants, and $k=2\pi/L$ represents the wavenumber. By neglecting the nonlinear terms in the kinematic condition \eqref{eq:KBC2} and the dynamic boundary condition \eqref{eq:Bern1}, the equations can be linearized as follows
	\begin{eqnarray}
		&&-\left(c + \gamma b\right) \hat{\eta_X} = \hat{\phi_Y},\label{eq:KBC2lin} \\
		&&\hat{V}_Y = -\hat{\eta}_X, \label{eq:KBCV2lin}\\
		&&-\left(c + \gamma b \right)\hat{\phi}_X +   \gamma \hat{\psi}  - E_b \hat{V}_X = 0. \label{eq:Bern2}
	\end{eqnarray}

By substituting equations (\ref{eq:linear2}) into equations (\ref{eq:KBC2lin})--(\ref{eq:Bern2}), it is revealed that the unknown constants $M$ and $N$ are given by
	\begin{align}
	M & = \frac{-iA(c+\gamma b)}{\sinh k }\,, \\
	N & = \frac{iA}{\sinh k }\,,
\end{align}
	respectively. Furthermore, the linear dispersion relation is obtained as
	\begin{equation}\label{eq:c}
		c_{\pm} = -\gamma  + \frac{1}{2k} \left[ \gamma \tanh k \pm \sqrt{\Delta} \right],
	\end{equation}
	where the discriminant $\Delta$ is defined as
	\begin{eqnarray}\label{eq:delta}
		\Delta = \gamma^2 \tanh^2 k +4k^2 E_b.
	\end{eqnarray}

In the reference frame moving with the speed of the wave, the trajectory of a fluid particle is described by $(X(t),Y(t))$, satisfying the equation
		\begin{equation}
	\begin{cases}
		\displaystyle\frac{dX}{dt} = \phi_X(X,Y) - \gamma(Y+1)-c \\[10pt]
			\displaystyle\dfrac{dY}{dt} = \phi_Y(X,Y). \\
			\end{cases}
		\end{equation}
		Under the linear theory, this system can be approximated as		
			\begin{equation}
				\begin{cases}
			\dfrac{dX}{dt} \approx - \gamma(Y+1)-c\\[10pt] 
			\dfrac{dY}{dt} \approx  - 0.\\
			\end{cases}
	\end{equation}

	Hence,  if  $Y^*$ denotes the location of a stagnation point we have that 
		 \begin{equation}
		 	 -\gamma(Y^*+1)-c = 0,
	 \end{equation}
		 where $c$ is given by the formula (\ref{eq:c}). 
		 
		 To ensure that the stagnation point lies within the physical domain, we impose the condition $Y^* \in [-1, 0]$. This condition is guaranteed if and only if
		 		\begin{equation} \label{eq:cond1}
		Y^* =  -\frac{\tanh(k)}{2k} - \sqrt{ \left(\frac{\tanh(k)}{2k}\right)^2 + \frac{E_b}{\gamma^2} }
		\end{equation}
		 and 
		\begin{equation}\label{eq:cond2}
			kE_b - \gamma^2 (k-\tanh(k)) <0.
		\end{equation} 
		
		Therefore, based on equations (\ref{eq:cond1}) and (\ref{eq:cond2}), we can draw the following conclusions:
		
		\begin{itemize}
			\item When $k$ and $\gamma$ are fixed, increasing the value of $E_b$ from zero to higher values leads to the destruction of stagnation points.

			\item When $\gamma$ and $E_b$ are fixed, in the shallow water limit ($k \approx 0$), the depth of the stagnation point is given by
			$$	Y^* =  -\frac{1}{2} - \sqrt{\left(\frac{1}{2}\right)^2 + \frac{E_b}{\gamma^2}} < -1.$$
			This implies that in the linear long wave regime, there are no stagnation points within the bulk of the fluid.
			
			\item When $\gamma$ and $E_b$ are fixed, in the deep water regime ($k\to \infty$), the depth of the stagnation point is given by
			$$ Y^* = -\sqrt{\frac{E_b}{\gamma^2}} \quad \text{and} \quad E_b < \gamma^2.$$
		\end{itemize}

	The aforementioned observations, although valid only for linear theory, provide valuable insights into the behaviour of  stagnation points in nonlinear waves, as will be elaborated upon in the next section.

	\section{Results from the Full Euler Equations}

	In this section, we explore solutions of the full Euler equations (\ref{eq:field1})-(\ref{eq:Bern1}) across  a range of wave amplitude, vorticity, and electric Bond number parameters.
	
	The amplitude parameter, denoted by $\varepsilon$, is defined as the wave steepness, given by the ratio of wave height $H$ to wavelength $L$. In our experiments, the depth is normalized to $1$ and we analyse  setups  of depth regime that  corresponds to  shallow,   intermediate  and deep water. Our main purpose  is to  investigate the impact of the intensity of the electric Bond number $E_b$  on the free surface, in the emergence of stagnation points and on the shape of  the level curves of the voltage potential.

			 \begin{figure}[!htb]
			 	\centering
						 	
			 	\includegraphics[scale=1]{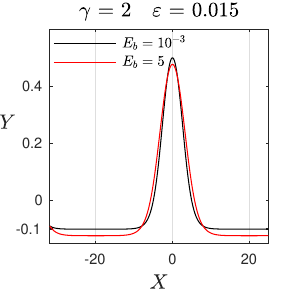}
			 	\includegraphics[scale=1]{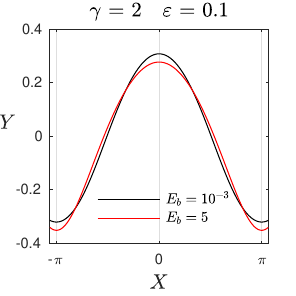}
			 	\includegraphics[scale=1]{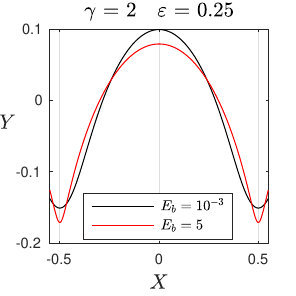}
			 	
			 	\includegraphics[scale=1]{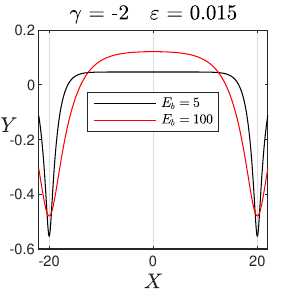}
			 	\includegraphics[scale=1]{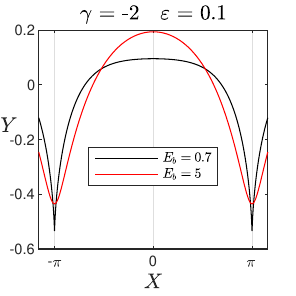}
			 	\includegraphics[scale=1]{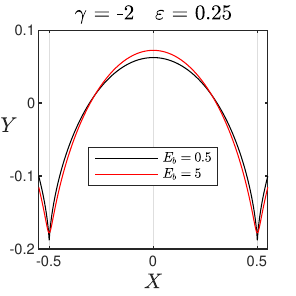}

			 	\caption{The effect of  the intensity of the electric field ($E_b$) on the shape of the free surface  wave in different depth regimes:  shallow water (left), intermediate water  (middle) and deep water (right).  The parameter $\varepsilon$ is  the steepness of the wave.}
			 	\label{fig:fig2}
			 \end{figure}

	\subsection{The effect of the electric field  on the  free surface}
	
	The effect of horizontal electric field on the free surface wave is a complex and active area of research. Among the facts that can influence the interaction between the electric field and the wave, we mention  the strength of the electric field and   the wavenumber. We refer to the work of   Papageorgiu \cite{Papageorgiu:2019} for a enlighting  review on experimental, numerical and mathematical studies on electrohydrodynamic flows.  In general, a horizontal electric field tend to stabilise  the wave.

	We  focus on investigating  how the intensity of the electric Bond number ($E_b$) affects wave profiles and wave speeds for a range of amplitude, wavelength and vorticity parameters. 	Figure \ref{fig:fig2} illustrates the influence of $E_b$ on the shape of waves in different water depth regimes: shallow water regime (left column), intermediate regime (center column), and deep water regime (right column). The simulations are performed for flows with vorticity values $\gamma =2,$ and $-2$. Here we do not present surface profiles for the irrotational regime $\gamma=0$, since in this case waves with arbitrary shape propagate along the liquid boundary without distortions.
Upon fixing the depth regime, we observe that waves in the shallow water regime are particularly sensitive to variations in the electric field, with this impact being more pronounced in flows with negative vorticity.
As the intensity of the electric field increases, it tends to gradually smooth the wave profile. Specifically, the sharp troughs noticeable at lower values of $E_b$ become more smooth with the increase of the intensity of the electric field. This phenomenon can be read as a kind of stabilization of the flow caused by the horizontal electric field. 	
	\begin{figure}[!ht]
		\centering
		\includegraphics[scale=1]{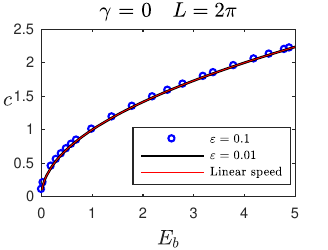}
		\includegraphics[scale=1]{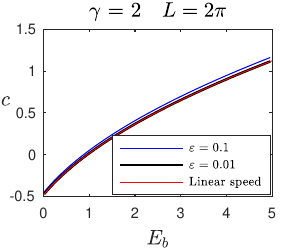}
		\includegraphics[scale=1]{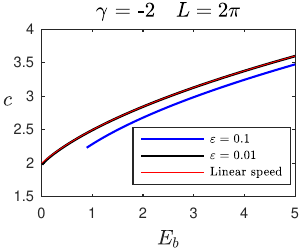}
		\includegraphics[scale=1]{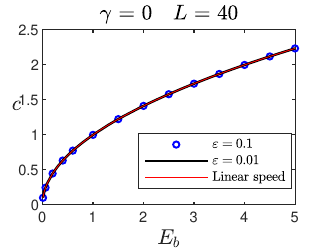}
		\includegraphics[scale=1]{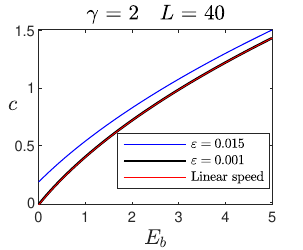}
		\includegraphics[scale=1]{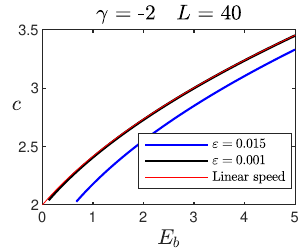}
		\includegraphics[scale=1]{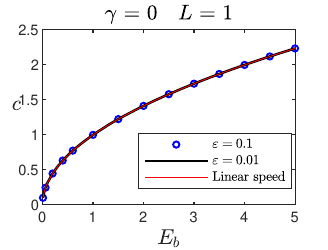}
		\includegraphics[scale=1]{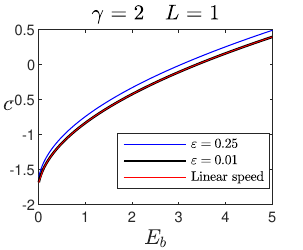}
		\includegraphics[scale=1]{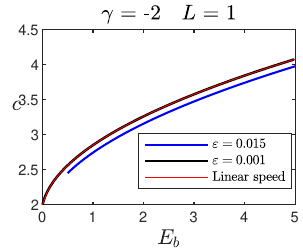}
		
		\caption{The wave speed $c$ as a function of $E_b$ for waves in a intermediate depth regime ($\lambda = 2\pi$). The parameter $\varepsilon$ is  the steepness of the wave.}
		\label{fig:fig3}
	\end{figure}

	Figure \ref{fig:fig3}   shows the relationship between wave speed $c$ and $E_b$ for flows with vorticity  $\gamma = 0,2, -2$ for intermediate  regime ($L= 2\pi$), shallow water ($L = 40$) and deep water ($L = 1$). 
	Notably, the speed for $\varepsilon = 0.01$ aligns well with  the linear speed which is  given  by the formula    (\ref{eq:c}). Additionally, we notice that the velocity increases as  $E_b$ increases.  We  note that when vorticity is zero, the velocity remains constant irrespective of the $\varepsilon$ value. This observation ties with the discussion in equation (\ref{eq:Free surface}), where we noticed that for irrotational flows  the wave speed remains the same regardless of the wave profile.

		\begin{figure}[!ht]
		\centering
		\includegraphics[scale=1]{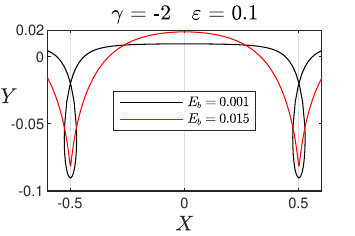}
		
		\caption{Effect of increasing  electric field intensity on the destruction of self-intersecting wave profiles.}
		\label{fig:fig4}
	\end{figure}

		In addition to the wave profiles shown in Figure~\ref{fig:fig2}, we observe self-intersecting waves, as depicted in Figure~\ref{fig:fig4}.  Our simulations suggest that higher intensities of the electric field tend to disrupt the self-intersecting wave profiles.   This observation is in agreement with the findings of Flamarion et al. \cite{Kochurin:2023}, who previously observed the spontaneous formation of self-intersecting wave profiles in flows with high vorticity and in the absence of an electric field. Their study highlighted  that the inclusion of a horizontal electric field  avoid the formation of such wave profile. 
		It is  important to point out that streamlines and stagnation points are only computed  beneath free surface  waves that do not self-intersect.

	\subsection{Streamlines and stagnation points}

	In this section, we investigate the influence of a horizontal electric field on the velocity field beneath periodic free surface waves. Specifically, we focus on analyzing the emergence and disappearance of stagnation points within the fluid bulk for different wave regimes.  To achieve this, we start by analyzing waves within the intermediate depth regime, where the wavelength is fixed at $L = 2\pi$. Subsequently, we  observe the transition between different regimes by varying the wavelength.	 It is important to say that  the phase portraits presented in this section are computed as level curves of the stream function (\ref{eq:streamfunction}) and  the precise identification of stagnation points is established by employing the velocity field formula in the canonical domain (see  \cite{Ribeiro:2017} for more details on that).

\begin{figure}[!ht]
		\centering
		\includegraphics[scale=1]{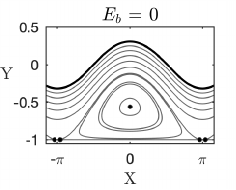}
		\includegraphics[scale=1]{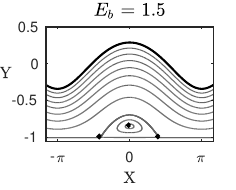}
			\includegraphics[scale=1]{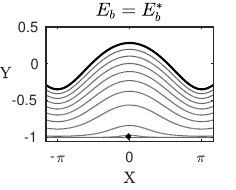}		
		\includegraphics[scale=1]{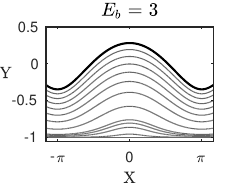}
			\caption{			
		Phase portrait for a flow with vorticity $\gamma = 2$ at varying electric Bond numbers, $E_b^* \approx 2.751 $. }
		\label{fig:fig5}
	\end{figure}

	In our first experiment, we keep the vorticity fixed at $\gamma = 2$ and the wave steepness at $\varepsilon = 0.1$, while varying the value of $E_b$. The resulting phase portrait, obtained by solving the full Euler equation, is presented in Figure \ref{fig:fig5}. Initially, three stagnation points are observed for $E_b = 0$. However, as the value of $E_b$ increases, the stagnation points gradually move closer,  coalesce at the bottom, and finally disappear. These findings suggest that a horizontal electric field acts as a mechanism for the elimination of stagnation points within the flow.  This behaviour  is the opposite  of the effect of  normal electric field where the increase of $E_b$ stimulates the  emergence of stagnation points. We refer to \cite{Flamarion:2022PoF} for further details  on normal electric fields.

\begin{figure}[!ht]
	\centering
	\includegraphics[scale=1]{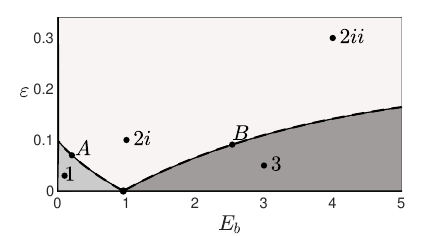}
	
	\includegraphics[scale=1]{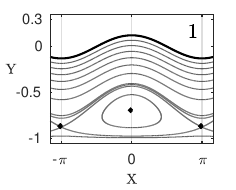}
\includegraphics[scale=1]{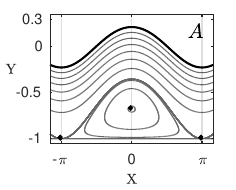}		
\includegraphics[scale = 1]{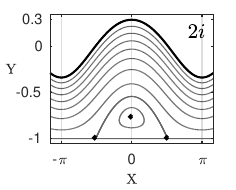}
\includegraphics[scale=1]{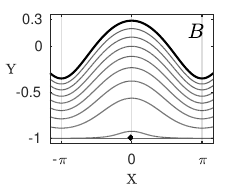}
\includegraphics[scale=1]{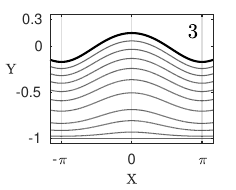}
\includegraphics[scale=1]{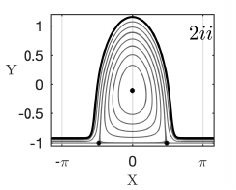}
	\caption{Variation in the flow structure as the wave steepness ($\varepsilon$) and  the intensity of the electric field ($E_b$) change, while maintaining fixed values for vorticity ($\gamma = 2$), depth ($1$), and wavelength ($L = 2\pi$).}
	\label{fig:fig6}
\end{figure}

A more comprehensive understanding of the bifurcations occurring in the parameter space $(E_b, \varepsilon)$ for $\gamma = 2$ is presented in Figure \ref{fig:fig6} (top). This parameter space is divided into three distinct regions:
\begin{description}
 \item [Region 1:] In this region, the flow exhibits two stagnation points within the fluid bulk: one centre beneath the crest and one saddle beneath the trough. A typical phase portrait for choices of $(E_b, \varepsilon)$ in this region is displayed in Figure \ref{fig:fig6} -1.

\item [Region 2:] Flows in this region possess three stagnation points: one centre  within the  bulk of the fluid  and two saddle points located at the bottom (see Figure 6 $2i$ and $2ii$)

\item [Region 3:] In this region, there are no stagnation points beneath the free surface wave.
\end{description}
When $\varepsilon = 0$, according to the linear theory formula (\ref{eq:cond1}), we find that the flow admits stagnation points at the bottom boundary beneath both the trough and crest of the wave when $E_b \approx 0.953$. Notably, the point $(0, 0.953)$ lies on the boundary of regions 1, 2, and 3.
A dashed line, marked as line A, separates regions 1 and 2. Along line A, there exist two stagnation points per wavelength: one centre beneath the crest and one saddle beneath the trough at the bottom -- see  Figure \ref{fig:fig6}-$A$.  Moving from region 1 to 2 along line A, the saddle points,  which are initially   one wavelenght apart, reach the bottom and subsequently  split into two other ones. 
On the dashed line B, a bifurcation occurs when crossing from region 2, with three stagnation points,  to region 3, with no stagnation in the fluid. On this bifurcation line, there exists a single stagnation point located at the bottom right under the crest -- see  Figure \ref{fig:fig6}-$B$. As the nonlinearity of the wave  is increased by elevating the values of $\varepsilon$, the free surface wave tends to assume a profile with almost flat troughs that nearly reach the bottom, along with a dominant recirculation zone within the flow (see Figure \ref{fig:fig6}-$2ii$). This feature indicates that, in this scenario, vorticity controls the flow. This observation leads to the conclusion that increasing nonlinearity results in vorticity dominating over the electric field.

\begin{figure}[!ht]
	\centering
	\includegraphics[scale= 1.00]{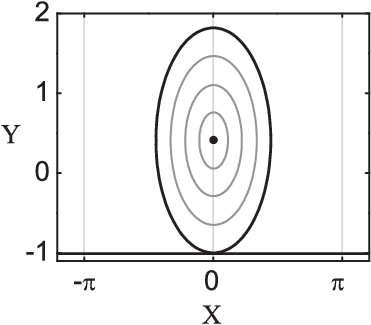}
	\caption{The limiting wave configuration corresponding to the maximum possible nonlinearity. The solution is a rigid rotation of the fluid around the stagnation point. }
	\label{fig:fignew}
\end{figure}

It is interesting to discuss the nature of the flow shown in the Figure~\ref{fig:fig6}-$2ii$. It can be assumed that in the absence of an electric field ($E_b=0$), the solution with the maximum possible nonlinearity is a circular cylinder with a radius of  $\sqrt{2}\approx 1.41$, rolling along the bottom: see Figure~\ref{fig:fignew}. The regime of fluid motion shown in Figure~\ref{fig:fignew} is not a result of numerical simulation but the exact analytical solution of the problem. For such a limiting solution, the $X$ and $Y$ velocity components are  $\{-\gamma Y/2, +\gamma X/2\}$, which corresponds to the rigid rotation of the fluid (in a coordinate system moving with the wave) around the stagnation point. Note that the flows shown in Figs.~\ref{fig:fig5} and~\ref{fig:fig6} can be interpreted as a nonlinear superposition of exact solutions of the problem, realized in three different limiting cases: (i) trivial solution   $\varepsilon=0$ ($\gamma$ and $E_b$  are arbitrary), (ii) solution with arbitrary $\varepsilon$  for $\gamma=0$ , and (iii) solution shown in Figure~\ref{fig:fignew} with the maximum possible steepness for  $E_b=0$. These three regimes of motion represent three limiting cases in which the problem has analytical solutions.

\begin{figure}[!ht]
	\centering
		\includegraphics[scale= 1]{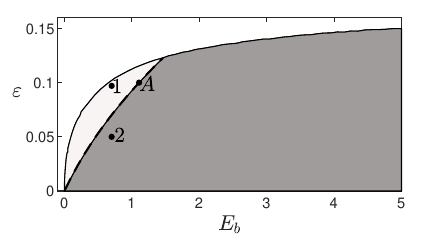}
		
	\includegraphics[scale= 1]{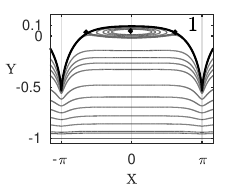}
	\includegraphics[scale= 1]{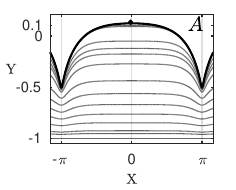}
	\includegraphics[scale= 1]{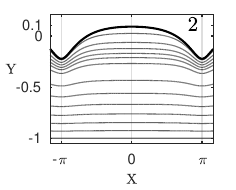}
	
\caption{Variation in the flow structure as the wave steepness ($\varepsilon$) and the intensity of the electric field ($E_b$) change, while maintaining fixed values for vorticity ($\gamma = -2$), depth ($1$), and wavelength ($L = 2\pi$).}
		\label{fig:fig7}
\end{figure}

In Figure \ref{fig:fig7}, we illustrate the parameter space $(E_b, \varepsilon)$ and provide typical phase portraits for flows with vorticity $\gamma = -2$. The dashed line  $A$ corresponds to bifurcation points at which the number of stagnation points changes. This line demarcates region $2$, which there is no  stagnation points in the fluid, from region $1$, which contains three stagnation points: two saddles at the free surface and a centre beneath the crest. Along line $A$, a single stagnation point is located at the wave crest. It is well-established that in the absence of the electric field, flows with negative vorticity do not have stagnation points (see \cite{Ribeiro:2017}). Therefore, the type of recirculation zone observed in Figure \ref{fig:fig7}-$1$ can be read as  consequence of the electric field. To the best of our knowledge, this configuration has not been reported in the literature before.

\begin{figure}[!ht]
	\centering
	\includegraphics[scale=1]{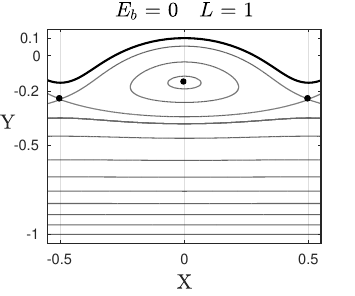}
	\includegraphics[scale=1]{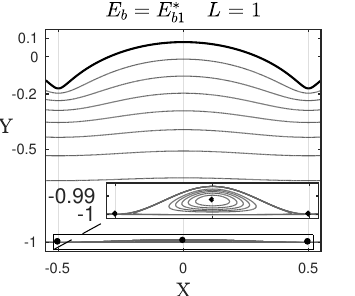}
        \includegraphics[scale=1]{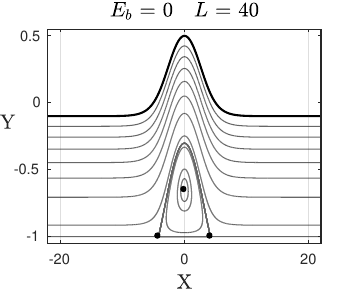}
	\caption{Typical phase portrait beneath waves in  deep and shallow water  for flows with vorticity  $\gamma = 2$. Here  $E_{b1}^* \approx 2.999 $}
	\label{fig:fig8}
\end{figure}

When we vary the depth regime from intermediate to  deep or shallow water, we observe that the electric field causes the same effect in the flow: the destruction of stagnation points. The primary difference between these  regimes and the intermediate regime is mainly characterized by the following: in deep water, waves with positive vorticity may feature saddle points closer to the trough of the wave, and when these saddle points reach the bottom, the resulting recirculation zone is notably narrower. In the shallow water regime, flows with positive vorticity only have saddle points at the bottom. Figure \ref{fig:fig8} depicts some typical phase portraits. It is noteworthy that these characteristics arise from the interplay between the wave regime and the intensity of vorticity, rather than being attributed to any discernible effects of the electric field. This observation aligns with analogous characteristics highlighted by Ribeiro-Jr et al. \cite{Ribeiro:2017} in their investigation of periodic waves within flows characterized solely by constant vorticity.

\subsection{Electric potential}
\begin{figure}[!ht]
	\centering
	\includegraphics[scale=1]{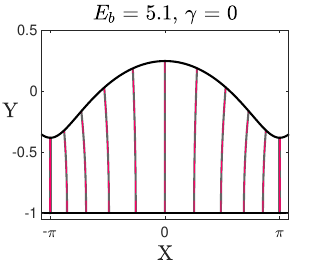}

	\includegraphics[scale=1]{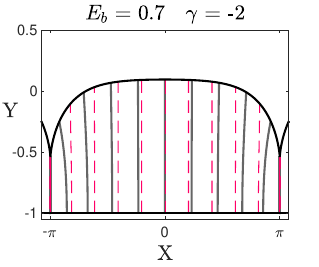}
	\includegraphics[scale=1]{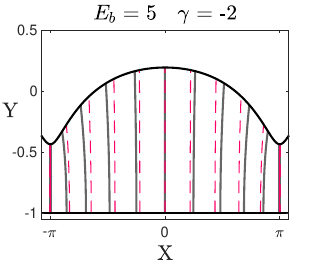}
	\includegraphics[scale=1]{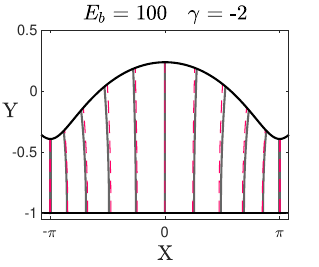}

	\includegraphics[scale=1]{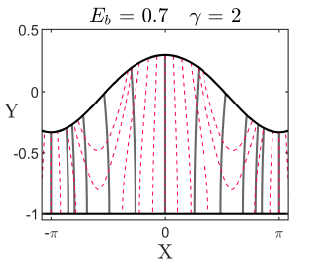}
	\includegraphics[scale=1]{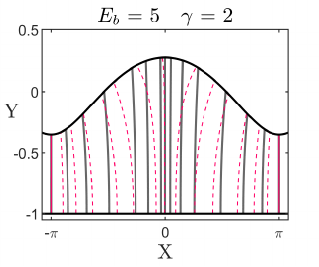}
		\includegraphics[scale=1]{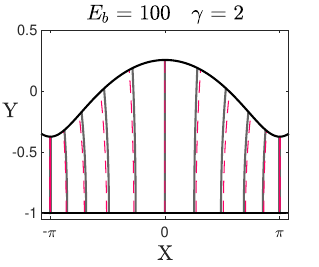}

	\caption{Level curves of the  voltage potential $V$(solid lines) and  of the quantity $\phi_T$(dashed  lines). Both potentials are normalized to one for comparative analysis. Parameter: $\varepsilon = 0.1$.}
	\label{fig:fig9}
\end{figure}

The aim of this section is to clarify the physical reason for the destruction of stagnation points in a strong horizontal electric field. As was mentioned earlier, in the irrotational case, the governing equation system has an exact analytical solution in the form of a stationary wave with arbitrary shape, for more details, see \cite{Zubarev:2004, Zubarev:2009}. In this regime of fluid motion, the electric field lines coincide with the stream lines inside the fluid (the velocity potential equals the electric field potential). Thus, the question arises: how are the velocity and electric field potentials related in the constant vorticity regime? To answer it, Figure~\ref{fig:fig9} shows the equipotential line (dashed red line)  along which the electric potential $V$ is constant  and the level curves (solid black lines)  of  the quantity  
$$ \phi_T(X,Y) \equiv \phi(X,Y) - c X.$$
When  the flow is irrotatinal this quantity $\phi_T(X,Y)$  corresponds to the velocity potential in the wave moving frame associated with the flow.
Remarkably,   for $\gamma = 0$, the curves appear nearly identical indicating  that $\phi_T \varpropto V$, which corresponds to  Zubarev's solution \cite{Zubarev:2004}. In the regime of a strong field but finite vorticity, the equipotentials for $V(X,Y)$ and $ \phi_T(X,Y)$ deviate slightly from each other, see Figure~\ref{fig:fig9}.Thus, as the electric field increases, the velocity potential tends to take the form of the electric potential. Since, the electric field in a liquid is potential (irrotational) it leads to the destruction of stagnation points.

\section{Discussion of the results and conclusions}

In this study, the examination focused on traveling periodic waves occurring on the surface of a finite depth layer composed of non-conducting fluid characterized by a high dielectric constant. 
  Our investigation concentrated on scenarios where the sole driving forces behind the fluid motion are the tangential electric field and a constant vorticity.
Comprehending the behaviour of periodic waves influenced solely by these specific driving forces, without the interference of other mechanisms contributes to fundamental knowledge in fluid dynamics. Furthermore, this understanding offers  insights for configuring laboratory experiments and comprehending the dynamics of real-world systems. Such insights are particularly relevant in scenarios where these forces operate in conjunction with other complex forces, providing a clearer understanding of their individual impacts and collective interactions.

The results obtained in this study pave the way for a comprehensive discussion on the role of the vorticity and a tangential electric field on the characteristics of  the flow beneath a periodic free surface wave.

   Our investigation was executed through a pseudo-spectral numerical method founded on a conformal mapping technique.  Upon observing simulations where the electric field was set to zero,  thus the only driving force is the vorticity, we notice that the emergence of  saddle   points within the bulk of the fluid was comparatively more feasible in this scenario in contrast to rotational-gravity flows. This contrast is evident when examining both linear theory and nonlinear solution.
In the context of linear theory, as described in equations (\ref{eq:cond1}) and (\ref{eq:cond2}), it is easy to see  that setting $E_b=0$ leads to the existence of stagnation points for any given value of vorticity. Additionally, the nonlinear simulation, as depicted in Figure \ref{fig:fig6} (top), illustrates well-defined saddle points within the fluid bulk for a vorticity value of $\gamma = 2$.  This observation stands in contrast to the findings presented by Ribeiro-Jr et al. \cite{Ribeiro:2017} (Figure 3 and 5), where configurations featuring similar characteristics in rotational-gravity flows were reported exclusively for considerably higher vorticity values.

Figure \ref{fig:fig8} and the  linear theory investigation reveal a shared characteristic between rotational-gravity flows and pure rotational flows: in the context of linear waves in shallow water regime, stagnation points beneath them are absent. Conversely, nonlinear waves in this depth regime demonstrate a phase portrait characterized by two saddle points at the bottom. This observation underscores the pivotal role of the depth regime in determining the presence or absence of stagnation points, as demonstrated in  others studies \cite{Flamarion:2022PoF, WaveMotion}.

In our analysis of the combined influence of vorticity and a horizontal electric field, a notable observation emerges: the increase of the intensity of the electric field leads to the elimination of stagnation points and tends to smoothen the wave profile, as visually depicted in figures \ref{fig:fig2}, \ref{fig:fig5}- \ref{fig:fig7}. 
  This  characteristic and the  fact that self-intersecting  are destroyed in the presence of high horizontal electric field  (see  Figure \ref{fig:fig4})  aligns with existing understanding that horizontal electric fields have a stabilizing effect on flow dynamics  \cite{Papageorgiu:2019}.



The influence of the electric field on smoothening the free surface is particularly noteworthy. In contrast to rotational-gravity flows, where negative vorticity typically results in waves with sharp crests (as depicted in \cite{Ribeiro:2017}, figure 13), the introduction of a horizontal electric field serves to attenuate the crest sharpness, concurrently generating stagnation points at the free surface (see Figure \ref{fig:fig7}).
  The physical reason for the destruction of stagnation points  by a strong electric field is that the velocity potential tends to take a form of the electric potential, as shown in Figure~\ref{fig:fig9}. Thus, the irrotational (potential) electric field leads to disappearing the stagnation points.

The identification of a stagnation point at the crest of a rounded wave presents an unexpected phenomenon. Typically, a wave featuring a stagnation point at its crest represents the highest irrotational wave characterised by a sharp crest with an angle of 120 degrees \cite{Fraenkel,Vanden-Broeck,Varvaruca}. Hence, the phase portraits  depicted in  Figure \ref{fig:fig7} is an unexpected occurrence in this context.

	\section*{Acknowledgments}
  The work of M.V.F. and R.R.Jr was supported in part by   National Council Scientific and Technological Development (CNPq) under  Chamada  CNPq/MCTI/N$^\circ$  10/2023-Universal. The work of E.K. on Section~2 is supported by Russian Science Foundation project No. 23-71-10012.

	\section*{Data Availability Statement}

	\end{document}